\newcommand{\CLASSINPUTtoptextmargin}{0.8in}
\newcommand{\CLASSINPUTbottomtextmargin}{0.98in}
\documentclass[conference]{IEEEtran}
\IEEEoverridecommandlockouts
\usepackage{cite}
\usepackage{amsmath,amssymb,amsfonts}
\usepackage{graphicx}
\usepackage{textcomp}
\usepackage{xcolor}
\usepackage{algorithmic,algorithm}
\def\BibTeX{{\rm B\kern-.05em{\sc i\kern-.025em b}\kern-.08em
    T\kern-.1667em\lower.7ex\hbox{E}\kern-.125emX}}

\DeclareMathOperator{\Tr}{\mathrm{Tr}}
\DeclareMathOperator{\Rank}{\mathrm{Rank}}
\DeclareMathOperator{\Diag}{\mathrm{Diag}}
\newcommand{\Nt}{{N_\mathrm{t}}}

\newcommand\relphantom[1]{\mathrel{\phantom{#1}}}
\allowdisplaybreaks

\newtheorem{lem}{Lemma}
\newtheorem{thm}{Theorem}

\renewcommand{\baselinestretch}{0.98}

\begin{document}

\title{Power-Efficient Resource Allocation for Multiuser MISO Systems via Intelligent Reflecting Surfaces
}
\author{\IEEEauthorblockN{Xianghao Yu\IEEEauthorrefmark{1},
		Dongfang Xu\IEEEauthorrefmark{1},  Derrick Wing Kwan Ng\IEEEauthorrefmark{2}, and Robert Schober\IEEEauthorrefmark{1}}
	\IEEEauthorblockA{\IEEEauthorrefmark{1}Friedrich-Alexander-Universit\"{a}t Erlangen-N\"{u}rnberg, Germany,
	\IEEEauthorrefmark{2}The University of New South Wales, Australia\\
	Email: \IEEEauthorrefmark{1}\{xianghao.yu, dongfang.xu, robert.schober\}@fau.de,  \IEEEauthorrefmark{2}w.k.ng@unsw.edu.au}}

\maketitle

\begin{abstract}
Intelligent reflecting surfaces (IRSs) are regarded as key enablers of next-generation wireless communications, due to their capability of customizing the wireless propagation environment. In this paper, we investigate power-efficient resource allocation for IRS-assisted multiuser multiple-input single-output (MISO) systems. 
To minimize the transmit power, both the beamforming vectors at the access point (AP) and phase shifts at the IRS are jointly optimized while taking into account the minimum required quality-of-service (QoS) of the users. To tackle the non-convexity of the formulated optimization problem, an inner approximation (IA) algorithm is developed. Unlike existing designs, which cannot guarantee local optimality, the proposed algorithm is guaranteed to converge to a Karush-Kuhn-Tucker (KKT) solution. Our simulation results show the effectiveness of the proposed algorithm compared to baseline schemes and reveal that deploying IRSs is more promising than leveraging multiple antennas at the AP in terms of energy efficiency. 
\end{abstract}


\section{Introduction}
There has been a growing interest in green wireless communications to reduce the power consumption of wireless networks over the past decade \cite{8014295}. Various technologies for green communications have been proposed including cloud radio access networks (C-RANs), energy harvesting, and cognitive radio (CR) networks \cite{7264986}. However, these existing approaches share two common disadvantages. First, the deployment of centralized baseband unit (BBU) pools in C-RANs, equipping energy harvesting transceivers, and the signaling overhead for spectrum sensing in CR inevitably cause additional power consumption. Second, while signal processing and energy resources can be improved with these approaches, the wireless channels are treated as a ``black box" which cannot be  controlled as would be desirable for green communications.

Thanks to the rapid evolution of radio frequency (RF) micro-electro-mechanical systems (MEMS), the integration of intelligent reflecting surfaces (IRSs) into wireless communication systems has been recently proposed \cite{di2019smart,xu2020resource}. In particular, with programmable reflecting elements, IRSs are able to provide reconfigurable reflections of the impinging wireless signals \cite{8466374}. This unique property creates the possibility of customizing favorable wireless propagation environments, which can be exploited to further reduce the power consumption of wireless systems. More importantly, typical IRSs consume no power for operation as the reflecting elements are implemented by \emph{passive} devices, e.g., dipoles and phase shifters \cite{8990007}.
Furthermore, IRSs can be fabricated as artificial thin films attached to existing infrastructures, such as the facades of buildings, which greatly reduces the implementation cost. To sum up, IRSs are promising candidates for power-efficient green wireless communications, and, more remarkably, are cost-effective devices with the ability to manipulate the radio propagation environment \cite{di2019smart}. 
Nevertheless, to further reduce the power consumption of IRS-assisted wireless systems, the IRSs have to be delicately designed and integrated with conventional communication techniques, such as the transmit beamforming at access points (APs).

There are several works on the design of green IRS-assisted  communication systems.
For instance, the energy efficiency was maximized in \cite{8741198}, where suboptimal zero-forcing beamforming was assumed at the AP. Hence, a significant performance loss
is expected as the joint design of the beamformers and reflecting elements was not considered. Besides, the transmit power minimization problem was investigated for multiuser multiple-input single-output (MISO) systems \cite{8811733}, Internet-of-Things (IoT) applications \cite{9013643}, and simultaneous wireless information and power transfer (SWIPT) systems \cite{xujie}. In \cite{8811733,9013643,xujie}, based on alternating minimization (AltMin) and semidefinite relaxation (SDR) methods, the total transmit power was minimized while taking into account the minimum required quality-of-service (QoS) of the users.
However, the combination of AltMin and SDR techniques does not guarantee the local optimality of the corresponding algorithms. In particular, the solutions generated by the Gaussian randomization process needed when applying SDR are not guaranteed to satisfy the QoS constraints and to monotonically decrease the transmit power during AltMin. 

In this paper, we study the power-efficient resource allocation design for IRS-assisted multiuser MISO systems. The IRS is assumed to be implemented by programmable phase shifters. We investigate the joint design of the beamforming vectors at the AP and the phase shifts at the IRS for minimization of the total transmit power while guaranteeing a minimum required signal-to-interference-plus-noise ratio (SINR) at each user. Instead of employing the AltMin and SDR approaches, the inner approximation (IA)  method is proposed for tackling the non-convexity of the formulated optimization problem. 
By convexifying the non-convex constraints, a sequence of approximating convex programs are solved in the IA algorithm.
The proposed IA algorithm is guaranteed to converge to a Karush-Kuhn-Tucker (KKT) solution of the original optimization problem, which is the main difference compared to  existing  algorithms that cannot guarantee local optimality \cite{8811733,9013643,xujie}. Our simulation results reveal that the proposed IA algorithm outperforms the state-of-the-art SDR-based AltMin algorithms in terms of transmit power consumption. In addition, the deployment of IRSs is shown to be more energy-efficient than equipping multiple antennas at the AP.

\emph{Notations:} In this paper, $\jmath=\sqrt{-1}$ denotes the imaginary unit of a complex number.
Vectors and matrices are denoted by boldface lower-case and  capital letters, respectively.
$\mathbb{C}^{m\times n}$ stands for the set of all $m\times n$ complex-valued matrices; 
$\mathbb{H}^{m}$ represents the set of all $m\times m$ Hermitian matrices; 
$\mathbf{1}_m$ denotes the $m\times1$ all-ones vector; $\mathbf{I}_m$ is the $m$-dimensional identity matrix.
$\mathbf{A}^H$ stands for the conjugate transpose of matrix $\mathbf{A}$.
The $\ell_2$-norm of vector $\mathbf{a}$ is denoted as $||\mathbf{a}||_2$. The spectral norm, nuclear norm, and Frobenius norm  of matrix $\mathbf{A}$ are represented as $\left\Vert\mathbf{A}\right\Vert_2$, $\left\Vert\mathbf{A}\right\Vert_*$, and $\left\Vert\mathbf{A}\right\Vert_F$, respectively.
$\mathrm{diag}(\mathbf{a})$ represents a diagonal matrix whose main diagonal elements are extracted from vector $\mathbf{a}$;
$\mathrm{Diag}(\mathbf{A})$ denotes  a vector whose elements are extracted from the main diagonal elements of matrix $\mathbf{A}$.
The eigenvector associated with the maximum eigenvalue of matrix $\mathbf{A}$ is denoted by  $\boldsymbol{\lambda}_{\max}(\mathbf{A})$.
$\Rank(\mathbf{A})$ and $\Tr(\mathbf{A})$  denote the rank  and trace of matrix $\mathbf{A}$;
$\mathbf{A}\succeq\mathbf{0}$ indicates that $\mathbf{A}$ is a positive semidefinite (PSD) matrix.
For a real-valued continuous function $f(\mathbf{A})$, $\nabla_\mathbf{A}f$ denotes the gradient of $f$ with respect to matrix $\mathbf{A}$. 
$\mathbb{E}[\cdot]$  and $\Re(\cdot)$  stand for statistical expectation and the real part of a complex number, respectively.
$\mathbf{A}^\star$ denotes the  optimal value of an optimization variable $\mathbf{A}$.

\section{System Model}
In this section, we first present the considered IRS-assisted multiuser MISO system and formulate the optimization problem. Then, we discuss the existing approach for solving the problem and its main limitations.
\subsection{IRS-Assisted System Model}
\begin{figure}[t]
	\centering\includegraphics[height=4.5cm]{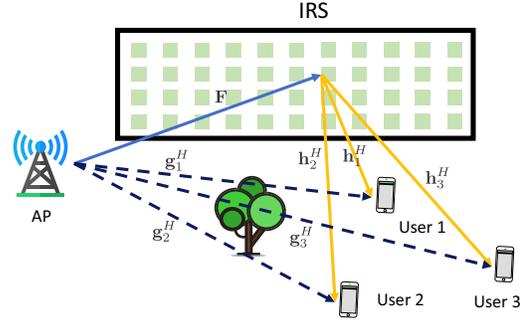}
	\caption{IRS-assisted multiuser MISO system consisting of $K=3$ users.}
	\label{model}
\end{figure}

We consider downlink transmission in an IRS-assisted multiuser MISO wireless communication system, which consists of an $\Nt$-antenna AP, $K$ single-antenna users, and an IRS, as shown in Fig. \ref{model}.
The IRS is implemented by $M$ programmable phase shifters.
The baseband signal received at user $k$ is given by
\begin{equation}
y_k=\left(\mathbf{h}_k^H\mathbf{\Phi}\mathbf{F}+\mathbf{g}_k^H\right)\sum_{j\in\mathcal{K}}\mathbf{w}_js_j+n_k,\quad\forall k\in\mathcal{K},
\end{equation}
where $\mathcal{K}=\{1,\cdots,K\}$. The IRS-user $k$ channel, AP-IRS channel, and AP-user $k$ channel are represented by $\mathbf{h}_k^H\in\mathbb{C}^{1\times M}$, $\mathbf{F}\in\mathbb{C}^{M\times\Nt}$, and $\mathbf{g}_k^H\in\mathbb{C}^{1\times\Nt}$, respectively. Since the IRS employs $M$ phase shifters, the phase shift matrix at the IRS is given by $\mathbf{\Phi}=\mathrm{diag}\left(e^{\jmath\theta_1},\cdots,e^{\jmath\theta_M}\right)$, where $\theta_m\in[0,2\pi]$, $\forall m\in\{1,\cdots,M\}$, represents the phase shift of the $m$-th reflecting element. The information-carrying signal transmitted to user $j$ is denoted by $s_j$, where $\mathbb{E}\left[|s_j|^2\right]=1$, $\forall j\in\mathcal{K}$, without loss of generality. The beamforming vector for user $j$ is denoted by $\mathbf{w}_j$. Variable $n_k$ represents the additive white Gaussian noise at user $k$ with zero mean and variance $\sigma_k^2$.
Therefore, the received SINR at user $k$ is given by
\begin{equation}\label{sinr}
\mathrm{SINR}_k=\frac{\left|\left(\mathbf{h}_k^H\mathbf{\Phi F}+\mathbf{g}_k^H\right)\mathbf{w}_k\right|^2}
{\sum_{j\in\mathcal{K}\backslash\{k\}}\left|\left(\mathbf{h}_k^H\mathbf{\Phi F}+\mathbf{g}_k^H\right)\mathbf{w}_j\right|^2+\sigma_k^2}.
\end{equation}

Our goal in this paper is to minimize the transmit power while ensuring a minimum required QoS of the users. The proposed power-efficient design of the beamformers and reflecting elements is obtained by solving the following optimization problem:
\begin{equation}
\begin{aligned}
&\underset{\mathbf{w}_k,\mathbf{\Phi}}{\mathrm{minimize}} && f\left(\mathbf{w}_k;\mathbf{\Phi}\right)=\sum_{k\in\mathcal{K}}\left\Vert\mathbf{w}_k\right\Vert_2^2\\
&\mathrm{subject\thinspace to}&&\mathrm{SINR}_k\ge\gamma_k,\quad\forall k,
\\
&&&\mathbf{\Phi}=\mathrm{diag}\left(
e^{\jmath\theta_1},e^{\jmath\theta_2},\cdots,e^{\jmath\theta_M}\right),
\end{aligned}\label{problem}
\end{equation}
where $\gamma_k$ is the predefined minimum required SINR of user $k$. 

\emph{Remark 1:} There are two main challenges in solving problem \eqref{problem}. First, each IRS reflecting element in $\mathbf{\Phi}$ has unit modulus, i.e., $\left|e^{\jmath\theta_m}\right|=1$, $\forall m\in\{1,\cdots,M\}$, which intrinsically is a non-convex constraint. Second, the optimization variables $\{\mathbf{w}_k\}_{k\in\mathcal{K}}$ and $\mathbf{\Phi}$ are coupled in the QoS constraint. These two facts make problem \eqref{problem} not jointly convex with respect to the optimization variables, and hence, in general difficult to  solve optimally.

\subsection{Existing Approach}
To tackle the difficulties in solving problem \eqref{problem}, SDR-based AltMin algorithms have been widely adopted in the literature \cite{8811733,9013643,xujie}. In particular,  the optimization of $\{\mathbf{w}_k\}_{k\in\mathcal{K}}$ and $\mathbf{\Phi}$ is decoupled and performed alternately by capitalizing on AltMin. For a given $\mathbf{\Phi}$, the optimization of the beamformers $\{\mathbf{w}_k\}_{k\in\mathcal{K}}$ is formulated as
\begin{equation}\label{p1}
\begin{aligned}
&\underset{\mathbf{w}_k}{\mathrm{minimize}} && f\left(\mathbf{w}_k\right)=\sum_{k\in\mathcal{K}}\left\Vert\mathbf{w}_k\right\Vert_2^2\\
&\mathrm{subject\thinspace to}&&\mathrm{SINR}_k\ge\gamma_k,\quad\forall k,
\end{aligned}
\end{equation}
which is identical to the corresponding problem in conventional wireless systems without IRSs, and therefore can be optimally solved via second-order cone programming (SOCP) \cite{8811733}. 
On the other hand, the phase shift matrix $\mathbf{\Phi}$ can be optimized by solving the following feasibility check problem:
\begin{equation}\label{eq4}
\begin{aligned}
&\underset{\mathbf{\Phi}}{\mathrm{minimize}} && 1\\
&\mathrm{subject\thinspace to}&&\mathrm{SINR}_k\ge\gamma_k,\quad\forall k,
\\
&&&\mathbf{\Phi}=\mathrm{diag}\left(
e^{\jmath\theta_1},e^{\jmath\theta_2},\cdots,e^{\jmath\theta_M}\right).
\end{aligned}
\end{equation}
According to \cite[eq. (44)]{8811733}, problem \eqref{eq4} can be reformulated as
\begin{equation}\label{eq5}
\begin{aligned}
&\underset{\mathbf{V}\in\mathbb{H}^{M+1}}{\mathrm{minimize}} && 1\\
&\mathrm{subject\thinspace to}&&\Tr\left(\mathbf{R}_k\mathbf{V}\right)\le\hat{\gamma}_k,\quad\forall k,
\\
&&&\Diag\left(\mathbf{V}\right)=\mathbf{1}_{M+1},\\
&&&\Rank\left(\mathbf{V}\right)=1,\\
&&&\mathbf{V}\succeq\mathbf{0},
\end{aligned}
\end{equation}
where $\hat{\gamma}_k=\left|\mathbf{g}_k^H\mathbf{w}_k\right|^2-\gamma_k\left(\sigma_k^2+\sum_{j\in\mathcal{K}\backslash\{k\}}\left|\mathbf{g}_k^H\mathbf{w}_j\right|^2\right)$, $|x|^2=1$, $\mathbf{v}=\left[e^{\jmath\theta_1},\cdots,e^{\jmath\theta_M},x\right]^H$, and $\mathbf{V}=\mathbf{vv}^H$. In addition, $\mathbf{R}_k=-\mathbf{T}_{k,k}+\gamma_k\sum_{j\in\mathcal{K}\backslash\{k\}}\mathbf{T}_{k,j}$, where $\mathbf{T}_{k,j}$ is given by $\mathbf{T}_{k,j}=$
\begin{equation}
\begin{bmatrix}
\mathrm{diag}\left(\mathbf{h}_k^H\right)\mathbf{F}\mathbf{w}_j\mathbf{w}_j^H\mathbf{F}^H\mathrm{diag}\left(\mathbf{h}_k\right)&\mathrm{diag}\left(\mathbf{h}_k^H\right)\mathbf{F}\mathbf{w}_j\mathbf{w}_j^H\mathbf{g}_k\\
\mathbf{g}_k^H\mathbf{w}_j\mathbf{w}_j^H\mathbf{F}^H\mathrm{diag}\left(\mathbf{h}_k\right)&0
\end{bmatrix}.
\end{equation}

One common approach to handle problem \eqref{eq5} is to first drop the non-convex rank-one constraint. The relaxed problem is then convex with respect to $\mathbf{V}$ and can be solved by standard convex program solvers such as CVX \cite{grant2008cvx}. 
Unfortunately, there is no guarantee that the obtained optimal solution $\mathbf{V}^\star$ is a rank-one matrix. A Gaussian randomization approach is therefore adopted to generate a unit modulus solution\footnote{Note that optimization variable $\mathbf{\Phi}$ in problems \eqref{problem} and \eqref{eq4} can be  determined once $\mathbf{v}$ has been obtained.} $\mathbf{v}$, i.e., $|v_i|=1$, $\forall i\in\{1,\cdots,M+1\}$, \cite{8811733,9013643,xujie}.
Although the optimal solution $\mathbf{V}^\star$ of the relaxed problem satisfies the QoS constraint, there is no guarantee that the randomized solution $\mathbf{v}$ also fulfills the constraint. In other words, the randomized solution $\mathbf{v}$ is not necessarily feasible for problem \eqref{eq4}. 

More importantly, in the  AltMin procedure, we have
\begin{equation}
f\left(\mathbf{w}_k^{(t)};\mathbf{v}^{(t)}\right)\overset{(a)}{=}f\left(\mathbf{w}_k^{(t)};\mathbf{v}^{(t+1)}\right)\overset{(b)}{\not\geq} f\left(\mathbf{w}_k^{(t+1)};\mathbf{v}^{(t+1)}\right),
\end{equation}
where $t$ is the iteration index. The equality in (a) is because problem \eqref{eq4} only finds a $\mathbf{v}$ that does not affect the objective value. On the other hand, the uncertainty in (b) means that the objective function does not necessarily decrease after solving problem $\eqref{p1}$ based on the $\mathbf{v}^{(t+1)}$ that has been obtained form problem \eqref{eq4}. 
In particular, for problem $\eqref{p1}$, $f\left(\mathbf{w}_k^{(t)};\mathbf{v}^{(t)}\right)$ and $f\left(\mathbf{w}_k^{(t+1)};\mathbf{v}^{(t+1)}\right)$ are optimal objective values based on the two different sets of parameters $\mathbf{v}^{(t)}$ and $\mathbf{v}^{(t+1)}$, respectively. 
When the parameter is updated from $\mathbf{v}^{(t)}$ to $\mathbf{v}^{(t+1)}$, the feasible set of problem \eqref{p1} changes. However, the relation between the two feasible sets cannot be quantified as both $\mathbf{v}^{(t)}$ and $\mathbf{v}^{(t+1)}$ may be infeasible for problem \eqref{eq4}.
Hence, there is no guarantee how the optimal objective value $f$ of problem \eqref{p1} improves when the parameter is updated from $\mathbf{v}^{(t)}$ to $\mathbf{v}^{(t+1)}$.

For the above mentioned two reasons, the state-of-the-art SDR-based AltMin algorithm is not guaranteed to converge to a locally optimal solution, as will be also verified in Section IV-A. This motives us to develop a novel algorithm design for solving problem \eqref{problem} in the next section.

\section{Design of IRS-Assisted Multiuser MISO Wireless Systems}
In the SDR-based AltMin algorithm, only one set of the optimization variables is updated in each iteration, such that local optimality cannot be guaranteed. Therefore, we develop an algorithm that optimizes all optimization variables in each iteration. To this end, we first reformulate problem \eqref{problem} as follows.
The numerator of the SINR in \eqref{sinr} is rewritten as
\begin{equation}
\begin{split}
&\left|\left(\mathbf{h}_k^H\mathbf{\Phi F}+\mathbf{g}_k^H\right)\mathbf{w}_k\right|^2=
2\Re\left[\tilde{\mathbf{v}}^H\mathrm{diag}\left(\mathbf{h}^H_k\right)\mathbf{FW}_k\mathbf{g}_k\right]\\
&+\tilde{\mathbf{v}}^H\mathrm{diag}\left(\mathbf{h}^H_k\right)\mathbf{FW}_k\mathbf{F}^H\mathrm{diag}\left(\mathbf{h}_k\right)\tilde{\mathbf{v}}+\mathbf{g}^H_k\mathbf{W}_k\mathbf{g}_k\\
&=\mathbf{v}^H\mathbf{G}_k^H\mathbf{W}_k\mathbf{G}_k\mathbf{v},
\end{split}
\end{equation}
where $\mathbf{G}_k=\begin{bmatrix}
\mathbf{F}^H\mathrm{diag}\left(\mathbf{h}_k\right)&\mathbf{g}_k
\end{bmatrix}$, $\mathbf{W}_k=\mathbf{w}_k\mathbf{w}_k^H$, and $\tilde{\mathbf{v}}=\left[
e^{\jmath\theta_1},e^{\jmath\theta_2},\cdots,e^{\jmath\theta_M}\right]^H$. Hence, the denominator can be rewritten in a similar manner and problem \eqref{problem} is reformulated as
\begin{align}
&\underset{\mathbf{W}_k\in\mathbb{H}^{\Nt},\mathbf{V}\in\mathbb{H}^{M+1}}{\mathrm{minimize}} && f(\mathbf{W}_k)=\sum_{k\in\mathcal{K}}\Tr\left(\mathbf{W}_k\right)\notag\\
&\mathrm{\quad\,\,\, subject\thinspace to}&&\mbox{C1:}\,\gamma_k\sigma_k^2+\gamma_k\sum_{k\in\mathcal{K}}\Tr\left(\mathbf{W}_j\mathbf{G}_k\mathbf{V}\mathbf{G}_k^H\right)\notag\\
&&&\relphantom{\mbox{C1}}-\Tr\left(\mathbf{W}_k\mathbf{G}_k\mathbf{V}\mathbf{G}_k^H\right)\le0,\quad\forall k,\notag\\
&&&\mbox{C2:}\,\mathrm{Diag}\left(\mathbf{V}\right)=\mathbf{1}_{M+1},\label{refor}\\
&&&\mbox{C3:}\,\mathrm{Rank}\left(\mathbf{V}\right)=1,\notag\\
&&&\mbox{C4:}\,\mathrm{Rank}\left(\mathbf{W}_k\right)=1,\quad\forall k,\notag\\
&&&\mbox{C5:}\,\mathbf{V}\succeq\mathbf{0},\quad\mbox{C6:}\,\mathbf{W}_k\succeq\mathbf{0},\quad\forall k.\notag
\end{align}
Next, we leverage the IA method to tackle the non-convex constraints $\mbox{C1}$ and $\mbox{C3}$ in problem \eqref{refor}. In particular, the general IA algorithm optimizes a sequence of approximating convex programs. In each iteration of the algorithm, the non-convex constraints are approximated by their convex counterparts.

\subsection{IA Method for QoS Constraint $\mbox{C1}$}
 We take the term $\Tr\left(\mathbf{W}_j\mathbf{G}_k\mathbf{V}\mathbf{G}_k^H\right)$ as an example to explain how we construct a convex constraint approximating the non-convex QoS constraint $\mbox{C1}$. The term is rewritten as
\begin{equation}\label{eq10}
\begin{split}
&\Tr\left(\mathbf{W}_j\mathbf{G}_k\mathbf{V}\mathbf{G}_k^H\right)=\frac{1}{2}\left\Vert\mathbf{W}_j+\mathbf{G}_k\mathbf{V}\mathbf{G}_k^H\right\Vert_F^2\\
&-\frac{1}{2}\Tr\left(\mathbf{W}_j^H\mathbf{W}_j\right)-\frac{1}{2}\Tr\left(\mathbf{G}_k\mathbf{V}^H\mathbf{G}_k^H\mathbf{G}_k\mathbf{V}\mathbf{G}_k^H\right).
\end{split}
\end{equation}
Now, constraint ${\mbox{C1}}$ can be rewritten in form of a difference of convex (d.c.) functions, where the last two terms in \eqref{eq10} are non-convex with respect to $\mathbf{W}_k$ and $\mathbf{V}$, respectively.
To facilitate IA, we  construct a global underestimator for the non-convex terms by first-order Taylor approximation. Specifically, we have 
\begin{align}
&\Tr\left(\mathbf{W}_j^H\mathbf{W}_j\right)\ge-\left\Vert\mathbf{W}_j^{(t)}\right\Vert_F^2+2\Tr\left(\left(\mathbf{W}_j^{(t)}\right)^H\mathbf{W}_j\right)\text{and}\notag\\
&\Tr\left(\mathbf{G}_k\mathbf{V}^H\mathbf{G}_k^H\mathbf{G}_k\mathbf{V}\mathbf{G}_k^H\right)\ge-\left\Vert\mathbf{G}_k\mathbf{V}^{(t)}\mathbf{G}_k^H\right\Vert_F^2\label{eq11}\\
&+2\Tr\left(\left(\mathbf{G}_k^H\mathbf{G}_k\mathbf{V}^{(t)}\mathbf{G}_k^H\mathbf{G}_k\right)^H\mathbf{V}\right),\notag
\end{align}
where
\begin{figure*}
	\begin{equation}\label{long}
		\begin{split}
			\overline{\mbox{C1}}\mbox{:}\,&\frac{1}{2}\left\Vert\mathbf{W}_k-\mathbf{G}_k\mathbf{V}\mathbf{G}_k^H\right\Vert_F^2+\frac{\gamma_k}{2}\sum_{j\in\mathcal{K}\backslash\{k\}}\left\Vert\mathbf{W}_j+\mathbf{G}_k\mathbf{V}\mathbf{G}_k^H\right\Vert_F^2-\gamma_k\sum_{j\in\mathcal{K}\backslash\{k\}}\Tr\left(\left(\mathbf{W}_j^{(t)}\right)^H\mathbf{W}_j\right)\\
			&-\Tr\left(\left(\mathbf{W}_k^{(t)}\right)^H\mathbf{W}_k\right)-\left[1+\gamma_k(K-1)\right]\Tr\left(\left(\mathbf{G}_k^H\mathbf{G}_k\mathbf{V}^{(t)}\mathbf{G}_k^H\mathbf{G}_k\right)^H\mathbf{V}\right)\\
			&+\gamma_k\sigma_k^2+\frac{1}{2}\left\Vert\mathbf{W}_k^{(t)}\right\Vert_F^2
			+\left[\frac{1}{2}+\frac{\gamma_k}{2}(K-1)\right]\left\Vert\mathbf{G}_k\mathbf{V}^{(t)}\mathbf{G}_k^H\right\Vert_F^2
			+\frac{\gamma_k}{2}\sum_{j\in\mathcal{K}\backslash\{k\}}\left\Vert\mathbf{W}_j^{(t)}\right\Vert_F^2\le0,\quad\forall k.
		\end{split}
	\end{equation}
	\hrule
\end{figure*}
 $\mathbf{W}_j^{(t)}$ and $\mathbf{V}^{(t)}$ are the solutions obtained in the $t$-th iteration, at which the Taylor expansions are performed.
In addition, the term $-\Tr\left(\mathbf{W}_k\mathbf{G}_k\mathbf{V}\mathbf{G}_k^H\right)$ in constraint $\mbox{C1}$ is upper bounded in a similar manner as $\eqref{eq10}$ and \eqref{eq11}, and therefore the non-convex constraint $\mbox{C1}$ is approximated by  constraint $\overline{\mbox{C1}}$ in \eqref{long}, shown at the top of this page.
Note that compared to constraint $\mbox{C1}$, where the optimization variables are coupled, the optimization variables are decoupled in constraint $\overline{\mbox{C1}}$, which is also jointly convex with respect to $\{\mathbf{W}_k\}_{k\in\mathcal{K}}$ and $\mathbf{V}$.

\subsection{IA Method for Rank-One Constraint $\mbox{C3}$}
Since it is difficult to directly derive an upper bound for the rank-one constraint $\mbox{C3}$, we first rewrite the rank-one constraint in equivalent form via the following lemma.
\begin{lem}
	The rank-one constraint $\mbox{C3}$ is equivalent to constraint $\widetilde{\mbox{C3}}$, given by
\end{lem}
\begin{equation}\label{force}
\widetilde{\mbox{C3}}\mbox{:}\,\left\Vert\mathbf{V}\right\Vert_*-\left\Vert\mathbf{V}\right\Vert_2\le0.
\end{equation}
\begin{IEEEproof}
	For any $\mathbf{X}\in\mathbb{H}^{m}$, the inequality $\left\Vert\mathbf{X}\right\Vert_*=\sum_i{\sigma_i}\ge\left\Vert\mathbf{X}\right\Vert_2=\underset{i}{\max}\{\sigma_i\}$ holds, where $\sigma_i$ is the $i$-th singular value of $\mathbf{X}$. Equality holds if and only if $\mathbf{X}$ has unit rank. 
\end{IEEEproof}
Now, constraint $\widetilde{\mbox{C3}}$ is written in form of d.c. functions. Therefore, by deriving the first-order Taylor approximation of $\left\Vert\mathbf{V}\right\Vert_2$ as 
\begin{equation}
\begin{split}
	\left\Vert\mathbf{V}\right\Vert_2&\ge
 \left\Vert\mathbf{V}^{(t)}\right\Vert_2+\Tr\Big[\boldsymbol{\lambda}_{\max}\left(\mathbf{V}^{(t)}\right)\\
 &\relphantom{\ge}\times\boldsymbol{\lambda}_{\max}^H\left(\mathbf{V}^{(t)}\right)\left(\mathbf{V}-\mathbf{V}^{(t)}\right)\Big],
\end{split}
\end{equation}
we obtain a convex approximation of constraint $\widetilde{\mbox{C3}}$, which is given by  constraint $\overline{\mbox{C3}}$ as follows:
\begin{equation}
\begin{split}
\overline{\mbox{C3}}\mbox{:}\,&\left\Vert\mathbf{V}\right\Vert_*-\Tr\Big[\boldsymbol{\lambda}_{\max}\left(\mathbf{V}^{(t)}\right)\boldsymbol{\lambda}_{\max}^H\left(\mathbf{V}^{(t)}\right)\\
&\times\left(\mathbf{V}-\mathbf{V}^{(t)}\right)\Big]-\left\Vert\mathbf{V}^{(t)}\right\Vert_2\le0.
\end{split}
\end{equation}
Therefore, a convex approximation $\overline{\mbox{C3}}$ of the non-convex rank-one constraint $\mbox{C3}$ is constructed and this constraint ensures that $\mbox{C3}$ is satisfied when the IA algorithm converges.

\subsection{Overall IA Algorithm}
With the approximated convex constraints $\overline{\mbox{C1}}$ and $\overline{\mbox{C3}}$ at hand, the optimization problem that has to be solved in the $(t+1)$-th iteration of the overall IA algorithm is given by
\begin{equation}\label{overall}
\begin{aligned}
&\underset{\mathbf{W}_k\in\mathbb{H}^{\Nt},\mathbf{V}\in\mathbb{H}^{M+1}}{\mathrm{minimize}} &&f(\mathbf{W}_k)= \sum_{k\in\mathcal{K}}\Tr\left(\mathbf{W}_k\right)\\
&\mathrm{\quad\,\,\, subject\thinspace to}&&\overline{\mbox{C1}},\mbox{C2},\overline{\mbox{C3}},\mbox{C4},\mbox{C5},\mbox{C6}.
\end{aligned}
\end{equation}
We note that the remaining non-convexity of problem \eqref{overall} stems from the $K$ rank-one constraints in \mbox{C4}. To tackle this issue, we remove constraint \mbox{C4} by applying SDR where the relaxed version of \eqref{overall} can be efficiently solved via standard convex program solvers such as CVX \cite{grant2008cvx}. The tightness of this SDR is revealed in the following theorem.
\begin{thm}
	An optimal beamforming matrix $\mathbf{W}_k$ satisfying $\Rank\left(\mathbf{W}_k\right)=1$ can always be obtained for problem \eqref{overall}.
\end{thm}
\begin{IEEEproof}
	Please refer to the Appendix.
\end{IEEEproof}
\begin{algorithm}[t]
	\caption{Inner Approximation (IA) Algorithm}
	\begin{algorithmic}[1]
		\STATE Initialize $\mathbf{V}^{(0)}$ with random phases and obtain $\mathbf{W}_k^{(0)}$ by solving problem $\eqref{p1}$. Set the convergence tolerance $\varepsilon$ and iteration index $t=0$;
		\REPEAT 
		\STATE For given $\mathbf{W}_k^{(t)}$ and $\mathbf{V}^{(t)}$, update $\mathbf{W}_k^{(t+1)}$ and $\mathbf{V}^{(t+1)}$ as the optimal solution of problem \eqref{overall} without $\mbox{C4}$;
		\STATE $t\leftarrow t+1$;
		\UNTIL $\frac{f\left(\mathbf{W}_k^{(t)}\right)-f\left(\mathbf{W}_k^{(t+1)}\right)}{f\left(\mathbf{W}_k^{(t+1)}\right)}\le\varepsilon$
	\end{algorithmic}
\end{algorithm}

The overall IA algorithm is summarized in \textbf{Algorithm 1}. According to \cite[Th. 1]{marks1978general}, the objective function $f$ in \eqref{refor} is non-increasing in each iteration and the proposed algorithm is guaranteed to converge to a KKT solution of problem \eqref{problem}.
The computational complexity of each iteration of the proposed IA algorithm is given by $\mathcal{O}\left(\log\frac{1}{\varepsilon}\left(K\Nt^{\frac{7}{2}}+M^{\frac{7}{2}}\right)\right)$, where $\mathcal{O}\left(\cdot\right)$ is the big-O notation \cite[Th. 3.12]{polik2010interior}.

\section{Simulation Results}
In this section, we evaluate the performance of the proposed IA algorithm. The system carrier center frequency is $2.4$ GHz while the noise power at each user is set to $\sigma_k^2=-90$ dBm, $\forall k$. The AP serves one sector of a cell with radius $R$, where $K$ users are randomly and uniformly distributed in this sector and the IRS is deployed at the edge of the cell.
The channel matrix $\mathbf{F}$ between  AP and IRS is modeled as
\begin{equation}
\mathbf{F}=\sqrt{L_0d^{-\alpha}}\left(\sqrt{\frac{\beta}{1+\beta}}\mathbf{F}^\mathrm{L}+\sqrt{\frac{1}{1+\beta}}\mathbf{F}^\mathrm{N}\right),
\end{equation}
where $L_0=\left(\frac{\lambda_{c}}{4\pi}\right)^2$ is a constant with $\lambda_{c}$ being the wavelength of the carrier frequency. The distance between AP and IRS is denoted by $d$ and $\alpha=2$ is the path loss exponent. The small-scale fading is assumed to be Ricean fading with Ricean factor $\beta=1$. $\mathbf{F}^\mathrm{L}$ and $\mathbf{F}^\mathrm{N}$ are the line-of-sight (LoS) and non-LoS components, respectively. The LoS component is the product of the receive and transmit array response vectors while the non-LoS component is modeled by Rayleigh fading. The channel vectors $\left\{\mathbf{h}_k\right\}_{k\in\mathcal{K}}$ are generated in a similar way as $\mathbf{F}$. In addition, the direct links $\left\{\mathbf{g}_k\right\}_{k\in\mathcal{K}}$ between AP and users are modeled as pure non-LoS channels, i.e., $\alpha=4$ and $\beta=0$, since one of the motivations for deploying IRSs is that the direct links are shadowed by obstacles. For the ease of presentation, we assume that the SINR thresholds for all users are identical $\gamma_k=\gamma$, $\forall k$. The number of random vectors generated by the Gaussian randomization in the existing SDR-based AltMin approach is $50$ and the convergence tolerance in the proposed IA algorithm is set to $\varepsilon=10^{-5}$.
\subsection{Convergence Performance}
\begin{figure}[t]
	\centering\includegraphics[height=5.6cm]{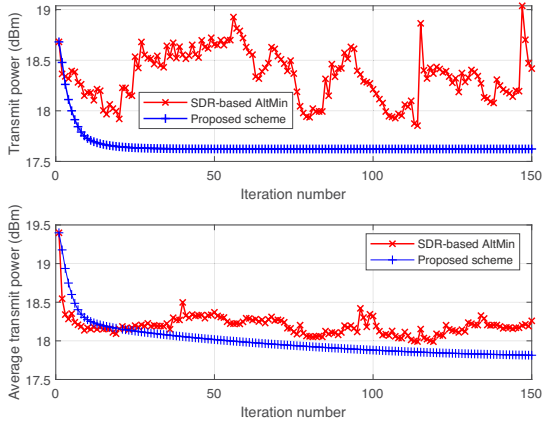}
	\caption{Convergence of different algorithms for $R=200$ m, $\Nt=M=10$, $K=3$, and $\gamma=2$ dB.} \label{fig0}
\end{figure}
The convergence of the SDR-based AltMin algorithm and the proposed IA algorithm is investigated for a typical snapshot and averaged over $500$ realizations in Fig. \ref{fig0}, respectively. As can be observed for the snapshot (upper half of Fig. \ref{fig0}), the objective function fluctuates significantly during AltMin, which confirms the analysis in Section II-B. In contrast, the proposed IA algorithm guarantees a monotonic convergence, which shows its superiority compared to the existing approach. While the oscillation is smoothed over a large number of realizations, the SDR-based AltMin algorithm still cannot guarantee the convergence of the average objective value, as shown in the lower half of Fig. \ref{fig0}. On the contrary, the proposed IA algorithm converges within 100 iterations on average. These results clearly show the motivation and importance of the proposed IA algorithm.

\subsection{Transmit Power Minimization}
\begin{figure}[t]
	\centering\includegraphics[height=5.6cm]{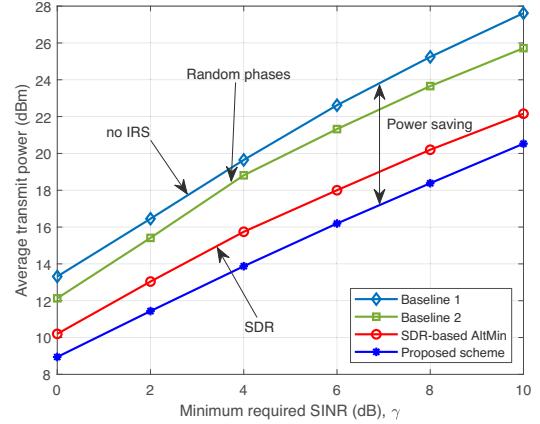}
	\caption{Average transmit power achieved by different algorithms when $R=100$ m, $\Nt=4$,  $M=20$, and $K=4$.} \label{fig1}
\end{figure}

In Fig. \ref{fig1}, the average transmit power at the AP is plotted for different algorithms. To show the effectiveness of the approach proposed in this paper, besides the SDR-based AltMin algorithm, two additional baseline schemes are considered.
For baseline scheme 1, we evaluate the transmit power when an IRS is not deployed.
For baseline scheme 2, we adopt an IRS implemented with random phases and optimize the beamformers by solving problem \eqref{p1}.
Since the SDR-based AltMin algorithm cannot guarantee convergence, for a fair comparison, we set the maximum iteration number equal to the number of iterations required by the IA algorithm to converge.
First, we observe that the required average transmit power is significantly reduced by deploying an IRS in the considered multiuser MISO system. 
This shows the ability of IRSs to establish favorable channel conditions, which facilitates achieving the QoS of the users at lower transmit powers. Hence, deploying IRSs is a promising approach for power-efficient wireless systems.
In addition, we note that the proposed IA algorithm outperforms both the SDR-based AltMin algorithm and the baseline scheme with random phases. This reveals the effectiveness of the proposed optimization methodology for jointly optimizing the beamformers and reflecting elements in IRS-assisted systems.

\subsection{Energy Efficiency Evaluation}
\begin{figure}[t]
	\centering\includegraphics[height=5.6cm]{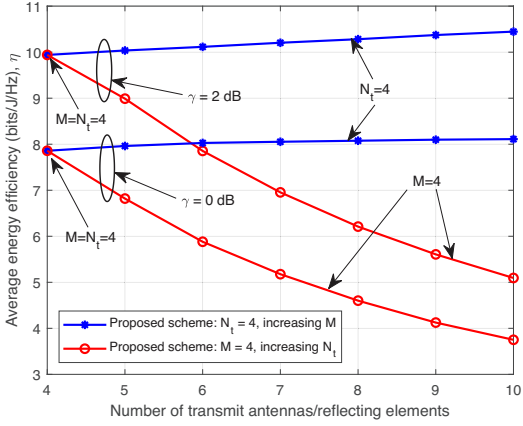}
	\caption{Average energy efficiency versus the number of transmit antennas, $\Nt$, or reflecting elements, $M$, when $R=100$ m, $K=4$, $\mu=0.32$, $P_\mathrm{s}=54$ mW, and $P_\mathrm{t}=100$ mW.}\label{fig2}
\end{figure}

IRSs are recognized as energy-efficient devices for improving communication performance.
In Fig. \ref{fig2}, we show the energy efficiency versus the number of antenna elements at the AP and the number of reflecting elements at the IRS. The energy efficiency is defined as \cite[eq. (32)]{my}
\begin{equation}
\eta=\frac{\sum_{k\in\mathcal{K}}\log_2\left(1+\mathrm{SINR}_k\right)}{\frac{1}{\mu}\sum_{k\in\mathcal{K}}\left\Vert\mathbf{w}_k\right\Vert_2^2+P_\mathrm{s}+\Nt P_\mathrm{t}},
\end{equation}
where $0<\mu\le1$ is the power amplifier efficiency, $P_\mathrm{s}$ is the static power consumed by the AP and IRS controller, and $P_\mathrm{t}$ accounts for the circuit power consumption introduced by deploying one antenna element.
We evaluate the average energy efficiency versus the number of reflecting elements for $\Nt=4$ transmit antennas (blue curves) and versus the number of transmit antennas for $M=4$ reflecting elements (red curves). As can be observed, the energy efficiency monotonically increases with the number of reflecting elements. In particular, additional reflecting elements  at the IRS provide more degrees of freedom for creating a  more favorable propagation environment which allows a further reduction of the transmit power. Moreover, deploying more reflecting elements does not consume additional power as they are passive devices. On the other hand, the energy efficiency of the system decreases as the number of transmit antennas equipped at the AP becomes large. 
This is because more circuit power is consumed if additional RF chains are deployed for driving the additional transmit antennas, which outweighs the transmit power reduction facilitated by employing more antennas. 
This observation strongly encourages the application of IRSs as power-efficient communication devices in  next-generation green wireless communication systems.

\section{Conclusions}
This paper studied the joint design of the beamforming vectors at the AP and the phase shifts at the IRS in an IRS-assisted multiuser MISO communication system. It was shown that the proposed IA algorithm is an effective design approach that effectively tackles the non-convexity of the formulated power minimization problem. Different from existing  algorithms that do not guarantee local optimality, one particular contribution of this paper is that the proposed IA algorithm is guaranteed to  converge to a KKT solution. 
Our simulation results revealed that IRSs have significant potential for the establishment of power-efficient green wireless communication systems.
\appendix
By relaxing the rank-one constraint $\mbox{C4}$ in problem \eqref{overall}, the remaining problem is jointly convex with respect to the optimization variables and satisfies Slater's constraint qualification.
Hence, strong duality holds and the Lagrangian function  is given by
\begin{align}\nonumber
{L}&=\sum_{k\in\mathcal{K}}\Tr\left(\mathbf{W}_k\right)+\sum_{k\in\mathcal{K}}\frac{\delta_k}{2}\left\Vert\mathbf{W}_k-\mathbf{G}_k\mathbf{V}\mathbf{G}_k^H\right\Vert_F^2\notag\\
&\relphantom{=}+\sum_{k\in\mathcal{K}}\frac{\delta_k\gamma_k}{2}\sum_{j\in\mathcal{K}\backslash\{k\}}\left\Vert\mathbf{W}_j+\mathbf{G}_k\mathbf{V}\mathbf{G}_k^H\right\Vert_F^2\notag\\
&\relphantom{=}-\sum_{k\in\mathcal{K}}\delta_k\gamma_k\sum_{j\in\mathcal{K}\backslash\{k\}}\Tr\left(\left(\mathbf{W}_j^{(t)}\right)^H\mathbf{W}_j\right)\\
&\relphantom{=}-\sum_{k\in\mathcal{K}}\delta_k\Tr\left(\left(\mathbf{W}_k^{(t)}\right)^H\mathbf{W}_k\right)-\sum_{k\in\mathcal{K}}\Tr\left(\mathbf{Y}_k\mathbf{W}_k\right)+\upsilon,\notag
\end{align}
where $\upsilon$ denotes the variables that are irrelevant to $\mathbf{W}_k$. $\delta_k\ge0$ and $\mathbf{Y}_k\in\mathbb{H}^\Nt$ are the Lagrange multipliers associated with constraints $\overline{\mbox{C1}}$ and $\mbox{C6}$, respectively. Then, we reveal the structure of $\mathbf{W}_k$ by examining the relevant KKT conditions of problem \eqref{overall}, which are given by  
\begin{equation}
\begin{split}
&\mbox{K1:}\,\delta_k^\star\ge0,\,\mathbf{Y}_k^\star\succeq\mathbf{0},\quad
\mbox{K2:}\,\mathbf{Y}_k^\star\mathbf{W}_k^\star=\mathbf{0},\\
&\mbox{K3:}\,\nabla_{\mathbf{W}_k}\mathcal{L}\left(\mathbf{W}_k^\star\right)=\mathbf{0}.
\end{split}
\end{equation}
With some basic algebraic manipulations, the KKT condition $\mbox{K3}$ can be rewritten as
\begin{equation}
\mathbf{Y}_k^\star=\mathbf{I}_\Nt-\boldsymbol{\Delta}_k^\star,
\end{equation}
where $\boldsymbol{\Delta}_k^\star=\delta_k^\star\mathbf{G}_k\mathbf{V}\mathbf{G}^H_k
-\sum_{j\in\mathcal{K}\backslash\{k\}}\delta_j^\star\gamma_j\mathbf{G}_j\mathbf{V}\mathbf{G}^H_j
+\left(\delta_k^\star+\sum_{j\in\mathcal{K}\backslash\{k\}}\delta_j^\star\gamma_j\right)\left(\mathbf{W}_k^{(t)}-\mathbf{W}_k^\star\right)$. By exploiting \cite[Appendix A]{xu2020resource}, it can be proved that $\Rank\left(\mathbf{Y}_k^\star\right)=\Nt-1$. According to KKT condition $\mbox{K2}$, this implies that we have $\Rank\left(\mathbf{W}_k\right)=1$, which completes the proof.

\bibliographystyle{IEEEtran}
\bibliography{conf}
\end{document}